\documentclass[11 pt]{article}
\usepackage[mathscr]{eucal}
\usepackage{amssymb,latexsym}
\usepackage{verbatim}
\usepackage{amsmath}
\usepackage{amsthm}
\usepackage{enumerate}
\usepackage{authblk}
\usepackage{color}
\usepackage{multicol}
\usepackage{tikz}
\usepackage{xypic}
\usepackage{caption}
\usepackage{cite}

\newtheorem{thm}{Theorem}[section]
\newtheorem{lem}[thm]{Lemma}
\newtheorem{cor}[thm]{Corollary}

\newcommand\wt{\widetilde}

\newcommand\s{\sigma}

\newcommand\e{\varepsilon}

\newcommand\g{\gamma}

\renewcommand\t{\tau}

\newcommand\beq{\begin{equation}}
\newcommand\eeq{\end{equation}}
\newcommand\ben{\begin{enumerate}}
\newcommand\een{\end{enumerate}}
\newcommand\bit{\begin{itemize}}
\newcommand\eit{\end{itemize}}

\newcommand{\R}{\mathbb R}

\newcommand{\ov}{\overline}
\newcommand{\ms}{\mathscr}
\newcommand{\half}{\frac{1}{2}}

\newcommand{\ext}{\text{{\rm ext}}}

\newcommand{\pd}{\partial}

\newcommand{\mc}{\mathcal}

\def\undertilde#1{\mathord{\vtop{\ialign{##\crcr
   $\hfil\displaystyle{#1}\hfil$\crcr\noalign{\kern1.5pt\nointerlineskip}
   $\hfil\tilde{}\hfil$\crcr\noalign{\kern1.5pt}}}}}

\newcounter{mnotecount}

\title{Remarks on the cosmological constant appearing as an initial condition for Milne-like spacetimes}
\author{Eric Ling\footnote{eling@math.rutgers.edu}}
\affil{Fields Institute
\\ University of Toronto}

\begin{document}
\date{}
\maketitle
\vspace{.2in}

\begin{abstract}
Milne-like spacetimes are a class of $k = -1$ FLRW spacetimes which admit continuous spacetime extensions through the big bang. In a previous paper \cite{Ling_coord_sing}, it was shown that the cosmological constant appears as an initial condition for Milne-like spacetimes. In this paper, we generalize this statement to spacetimes which share similar geometrical properties with Milne-like spacetimes but without the strong spatially isotropic assumption associated with them. We show how our results yield a ``quasi de Sitter" expansion for the early universe which could have applications to inflationary scenarios.
\end{abstract}


\vspace{.15in}

\section{Introduction}

Milne-like spacetimes are a class of $k = -1$ FLRW spacetimes which admit continuous spacetime extensions through the big bang. This extension was  observed in \cite{GalLing_con},\footnote{These extensions have been noted previously in the physics literature, see e.g. \cite{Coleman}.} and further properties of these spacetimes were explored in \cite{Ling_coord_sing}. Similar to how investigating the geometrical properties of the $r = 2m$ event horizon in the Schwarzschild spacetime led to a better understanding of black holes, we believe that investigating the geometrical properties of the big bang extension for Milne-like spacetimes may lead to a better understanding of cosmology.

In \cite[Thm. 4.2]{Ling_coord_sing}, it was shown that, under suitable hypotheses of the scale factor for a Milne-like spacetime, the equation of state for the energy density $\rho$ and pressure $p$ at the big bang is the same as that of a cosmological constant, namely, $\rho(0) = -p(0)$. We referred to this property as ``the cosmological constant appearing as an initial condition for Milne-like spcetimes." In this paper we generalize this statement to spacetimes which share similar geometrical properties with Milne-like spacetimes but without any homogeneous or isotropic assumptions. (Recall that Milne-like spacetimes are a subclass of FLRW spacetimes and hence are spatially isotropic.)

\newpage

De Sitter space is characterized by an equation of state $\rho = - p$ \cite[p. 124]{HE}. In this way, the initial equation of state $\rho(0) = -p(0)$ for Milne-like spacetimes can be described as yielding a ``quasi de Sitter" expansion for cosmic times $\tau$ close to the big bang $\tau = 0$, i.e. if $\rho(0) = -p(0)$, then $\rho(\tau) \approx -p(\tau)$ for $\tau$ near $\tau = 0$. This has applications to inflationary theory since $\rho(\tau) \approx -p(\tau)$ yields an inflationary era, $a''(\tau) > 0$, provided $\rho(0) > 0$.

This paper is divided as follows. In section \ref{Milne-like ext sec}, we review the definition of Milne-like spacetimes and their continuous spacetime extensions through the big bang. In section \ref{Milne-like cosmo const sec}, we review how the cosmological constant appears as an initial condition for Milne-like spacetimes. In section \ref{main result}, we prove our main results which generalize the results in section \ref{Milne-like cosmo const sec} to spacetimes without homogeneous or isotropic assumptions. Lastly, in section \ref{inflationary section}, we show how the ``quasi de Sitter" nature of these spacetimes can imply inflationary scenarios.

Our main result, Theorem \ref{main}, says that, under certain assumptions on a spacetime $(M,g)$, the cosmological constant appears as an initial condition. The point of the assumptions in the theorem is to remove the spatial isotropy enjoyed by Milne-like spacetimes. However, we have not been able to construct examples of such spacetimes that are not Milne-like. This leads to the following open question: are there spacetimes $(M,g)$ which satisfy the hypotheses of Theorem \ref{main} and are not Milne-like? To simplify matters, it's worth asking the same question but replacing assumption (a) in Theorem \ref{main} with ``$(M,g)$ solves the vacuum Einstein equations with a cosmological constant, i.e. $\text{Ric} = \Lambda g$."

Milne-like spacetimes were found by investigating low regularity aspects of Lorentzian geometry. This is a growing field with many tantalizing problems to solve. For low regularity causal theory, generalizations, and various results, see \cite{ChrusGrant, Leonardo, Ling_causal_theory, Minguzzi_cone, future_not_open, Clemens_GH, Lesourd_Minguzzi}. For low regularity spacetime inextendibility results, see \cite{SbierskiSchwarz1, SbierskiSchwarz2, SbierskiHol, GalLing_con, GLS, GrafLing, ChrusKlinger}. For the singularity theorems in low regularity, see \cite{Hawking_Penrose_C11, Hawking_sing_low_reg, Penrose_sing_low_reg, Graf_sing_thm, Schin_Stein}. For results on geodesics and maximizing causal curves in low regularity, see \cite{Clemens_Steinbauer, Lorentz_meets_Lipschitz, Schin_Stein}. For results on Lorentzian length spaces, see \cite{Lorentzian_length_spaces, cones_as_length_spaces, length_spaces_causal_hierarchy, time_fun_on_length_spaces, Lorentzian_analogue}. Lastly, for results related to the null distance function and other notions of distance defined on a spacetime, see \cite{Null_distance, Spacetime_distances_exploration, prop_null_dist, null_distance_lorentzian_length_spaces}. 

\medskip

\subsection{Milne-like spacetimes and their continuous spacetime extensions through the big bang}\label{Milne-like ext sec}

In this section, we review the definition of Milne-like spacetimes and their continuous spacetime extensions through the big bang.

\emph{Milne-like spacetimes} are $k = -1$ FLRW spacetimes satisfyng the following limiting condition on the scale factor: $a(\tau) = \tau + o(\tau^{1+\e})$ as $\tau \to 0$ for some $\e > 0$. Specifically, the manifold and metric are given by 
\begin{equation}
M \,=\, (0, \tau_{\rm max}) \times \R^3\:\:\:\: \text{ and } \:\:\:\: g \,=\, -d\tau^2 + a^2(\tau) h
\end{equation}
where $(\R^3, h)$ is hyperbolic space with constant sectional curvature $k = -1$. We assume $a(\tau)$ is smooth so that $(M,g)$ is a smooth spacetime. The \emph{Milne universe} corresponds to the scale factor $a(\tau) = \tau$  which isometrically embeds into Minkowski space.

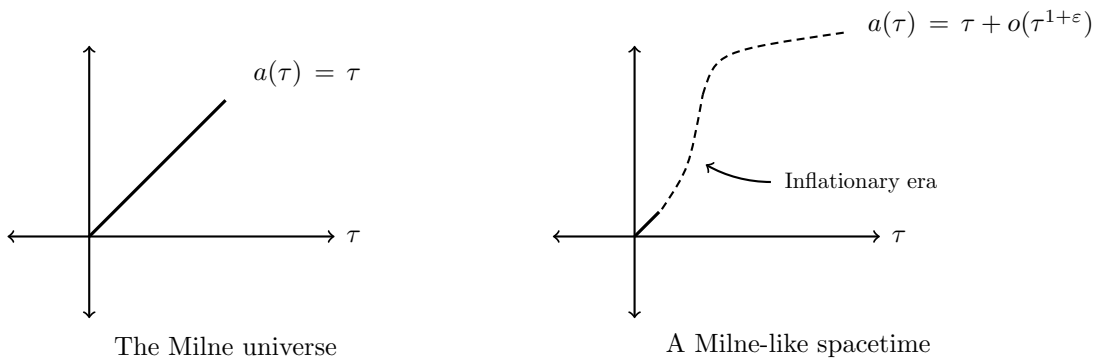
\begin{figure}[h]
\begin{center}
\begin{tikzpicture}[scale = .725]

\draw [<->,thick] (-12,-2.5) -- (-12,2.5);
\draw [<->,thick] (-13.5,-1) -- (-7.5,-1);
\draw [very thick] (-12,-1) -- (-9.5,1.5);

\draw (-7.15, -1) node {\small{$\tau$}};

\draw  (-8, 2) node {\small{$a(\tau) \,=\, \tau$}};

\draw (-9.5,-3) node {\small{The Milne universe}};


\draw [<->,thick] (-2,-2.5) -- (-2,2.5);
\draw [<->,thick] (-3.5,-1) -- (2.5,-1);

\draw (2.85, -1) node {\small{$\tau$}};

\draw [very thick] (-2,-1) -- (-1.55,-0.55);
\draw [densely dashed, thick] (-1.5, -0.5) .. controls (-1,.25).. (-.75,1.6);
\draw [densely dashed, thick] (-.75, 1.6) .. controls (-.5,2.4).. (1.9,2.75);

\draw  (4.35, 2.9) node {\small{$a(\tau) \,=\, \tau + o(\tau^{1 +\e})$}};

\draw [->] [thick] (0.5,0) arc [start angle=-90, end angle=-120, radius=68pt];
\draw (2.15,0) node [scale = .85]{\small{Inflationary era}};

\draw (1.0,-3) node {\small{A Milne-like spacetime}};

\end{tikzpicture}
\end{center}
\captionsetup{format=hang}
\caption{\small{Left: The scale factor for the Milne universe. Right: The scale factor for a Milne-like spacetime modeling an inflationary era.}}\label{milne universe and milne-like scale factor figure}
\end{figure}

Since the assumption on the scale factor is a limiting condition, Milne-like spacetimes can include an inflationary era, a radiation-dominated era, a matter-dominated, and a dark energy-dominated era. Hence they can model the dynamics of our universe. Figure \ref{milne universe and milne-like scale factor figure} depicts a Milne-like spacetime modeling an inflationary era. 

Introducing coordinates $(R, \theta, \phi)$ for the hyperbolic metric $h$, we can write the spacetime metric as
\begin{equation}
g \,=\, -d\tau^2 + a^2(\tau)\big[dR^2 + \sinh^2(R)(d\theta^2 + \sin^2\theta d\phi^2) \big].
\end{equation}
We introduce new coordinates $(t,r,\theta, \phi)$ via 
\begin{equation}\label{t and r def}
t \,=\, b(\tau)\cosh(R) \quad \text{ and } \quad r\,=\, b(\tau)\sinh(R),
\end{equation}
 where $b$ is given by $b(\tau) = \exp(\int_{\tau_0}^\tau \frac{1}{a(s)}ds)$ for some $\tau_0 > 0$. (Note that for the Milne universe, $a(\tau) = \tau$, we obtain $b(\tau) = \tau$ when $\tau_0 = 1$.) Hence $b$ satisfies $b' = b/a$. Putting $\Omega = 1/b' = a/b$, the metric in these new coordinates is
\begin{align}\label{conformal metric intro eq}
g \,&=\, \Omega^2(\tau)\big[-dt^2 + dr^2 + r^2(d\theta^2 + \sin^2\theta d\phi^2) \big] \nonumber
\\
&=\, \Omega^2(\tau)[-dt^2 + dx^2 + dy^2 + dz^2] \nonumber
\\
&=\, \Omega^2(\tau)\eta.
\end{align} 

\begin{figure}[h]
\begin{center}
\begin{tikzpicture}[scale = .7]

\shadedraw [white](-4,2) -- (0,-2) -- (4,2);
\shadedraw [dashed, thick, blue](0,-2) -- (4,2);
\shadedraw [dashed, thick, blue](0,-2) -- (-4,2);

\draw [<->,thick] (0,-3.5) -- (0,2.25);
\draw [<->,thick] (-4.5,-2) -- (4.5,-2);

\draw (-.35,2.5) node [scale = .85] {$t$};
\draw (4.75, -2.25) node [scale = .85] {$x^i$};
\draw (-.25,-2.25) node [scale = .85] {$\mc{O}$};

\draw [->] [thick] (1.5,2.8) arc [start angle=140, end angle=180, radius=60pt];
\draw (2.0,3.25) node [scale = .85]{\small{The Milne universe}};

\draw [->] [thick] (-2.4,-1.75) arc [start angle=-90, end angle=-30, radius=40pt];
\draw (-3.4,-1.7) node [scale = .85] {\small lightcone};


\draw [thick, red] (-3.84,2) .. controls (0,-2) .. (3.84,2);
\draw [thick, red] (-3.5,2) .. controls (0, -1.3).. (3.5,2);

\draw [->] [thick]  (1,-2.3) arc [start angle=-120, end angle=-180, radius=40pt];
\draw (2.3,-2.5) node [scale = .85] {\small{$\tau =$ constant }};

\draw (0,-4.5) node [scale = 1] {\small{$g \,=\, -dt^2 + dx^2 + dy^2 + dz^2$}};


\shadedraw [dashed, thick, white](9,2) -- (13,-2) -- (17,2);
\shadedraw [dashed, thick, blue](13,-2) -- (17,2);
\shadedraw [dashed, thick, blue](13,-2) -- (9,2);

\draw [<->,thick] (13,-3.5) -- (13,2.25);
\draw [<->,thick] (8.5,-2) -- (17.5,-2);

\draw (12.65,2.5) node [scale = .85] {$t$};
\draw (17.75, -2.25) node [scale = .85] {$x^i$};
\draw (12.75,-2.25) node [scale = .85] {$\mc{O}$};


\draw [->] [thick] (14.5,2.8) arc [start angle=140, end angle=180, radius=60pt];
\draw (15.0,3.25) node [scale = .85]{\small{A Milne-like spacetime}};

\draw [->] [thick] (10.6,-1.75) arc [start angle=-90, end angle=-30, radius=40pt];
\draw (9.6,-1.7) node [scale = .85] {\small lightcone};


\draw [thick, red] (9.16,2) .. controls (13,-2) .. (16.84,2);
\draw [thick, red] (9.5,2) .. controls (13, -1.3).. (16.5,2);

\draw [->] [thick]  (14,-2.3) arc [start angle=-120, end angle=-180, radius=40pt];
\draw (15.3,-2.5) node [scale = .85] {\small{$\tau =$ constant }};

\draw (13,-4.5) node [scale = 1] {\small{$g \,=\,\Omega^2(\tau)[ -dt^2 + dx^2 + dy^2 + dz^2]$}};

\end{tikzpicture}
\end{center}
\captionsetup{format=hang}
\caption{\small{Left: the Milne universe sits inside the future lightcone at the origin $\mc{O}$ of Minkowsi space. Right: a Milne-like spacetime sits inside the future lightcone at the origin $\mc{O}$ of a spacetime conformal to Minkowski space. In both cases the spacetime is foliated by the hyperboloids of constant $\t$ and extends continuously through the lightcone at $\mc{O}$.}}\label{milne universe and milne-like figure}
\end{figure}
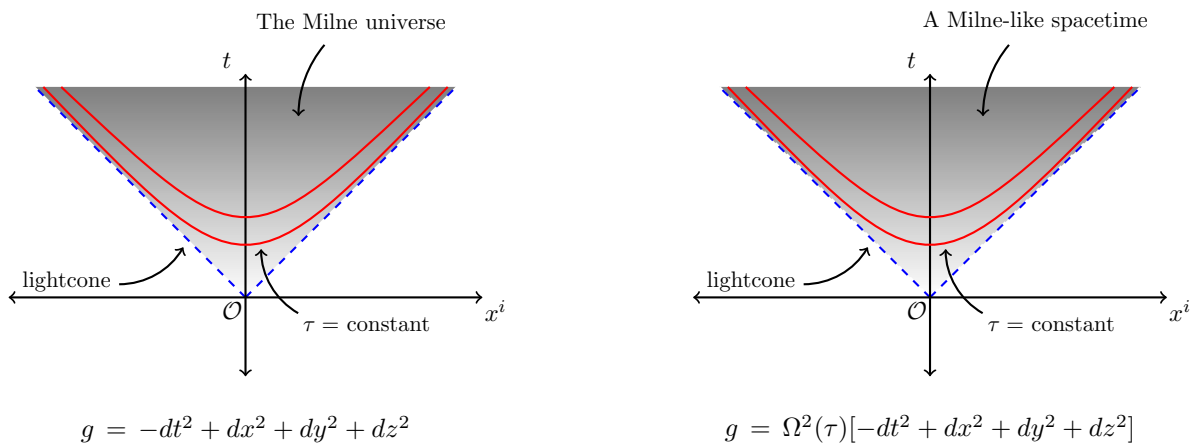

Thus Milne-like spacetimes are conformal to (a subset of) Minkowski space. In eq. (\ref{conformal metric intro eq}), $\tau$ is implicitly a function of $t$ and $r$. Specifically, $\tau$ is related to $t$ and $r$ via 
\begin{equation}\label{tau t r eq}
b^2(\tau) \,=\, t^2 - r^2. 
\end{equation}
Therefore the spacetime manifold $M$ lies within the set of points $t^2 - r^2 > 0$. Since $t > 0$ by eq. (\ref{t and r def}), it follows that $M$ lies within the set of points $t > r$. See figure \ref{milne universe and milne-like figure}.

The proof of \cite[Thm. 3.4]{Ling_coord_sing} shows that $b(0) = 0$ where $b(0) = \lim_{\tau \to 0}b(\tau)$. Therefore, by eq. (\ref{tau t r eq}), $\tau = 0$ corresponds to the set of points $t = r$ on the lightcone at the origin $\mc{O}$. Lastly, the proof also shows that $\Omega(0) = \tau_0$. Since $\tau_0 > 0$,  eq. (\ref{conformal metric intro eq}) implies that there is no degeneracy at $\tau = 0$ in these coordinates (i.e. the big bang is a coordinate singularity for Milne-like spacetimes). Therefore Milne-like spacetimes admit continuous spacetime extensions through the big bang by defining the extended metric $g_\ext$ via $g_\ext = \Omega^2(0)\eta$ for points $t \leq r$ and $g_\ext = g$ for points $t > r$. ``Continuity" here refers to the fact that the metric $g_\ext$ is merely continuous.\footnote{Using similar arguments as in \cite[Appendix B]{Greg_Graf_Ling_AdSxS2}, one can show that Milne-like spacetimes actually admit \emph{Lipschitz} spacetime extensions through the big bang. This should be compared with the results in \cite{SbierskiHol}.}

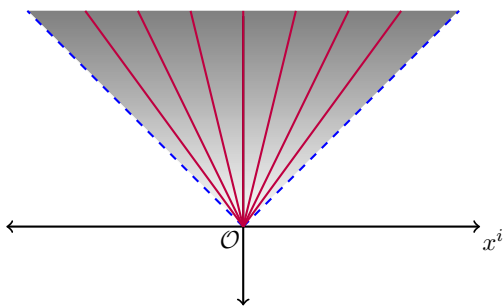
\begin{figure}[h]
\begin{center}
\begin{tikzpicture}[scale = 0.7]

\shadedraw [white] (-4.1,2.1) -- (0,-2) -- (4.1,2.1);
\draw [dashed, thick, blue] (0,-2) -- (4.1,2.1);
\draw [dashed, thick, blue] (0,-2) -- (-4.1,2.1);

\draw [<-,thick] (0,-3.5) -- (0,2.0);
\draw [<->,thick] (-4.5,-2) -- (4.5,-2);

\draw (4.75, -2.25) node [scale = .85] {$x^i$};
\draw (-.25,-2.25) node [scale = .85] {$\mc{O}$};
%

	\draw [thick, purple] (0,-2) -- (2,2.1);
	\draw [thick, purple] (0,-2) -- (3,2.1);
	\draw [thick, purple] (0,-2) -- (1,2.1);
	\draw [thick, purple] (0,-2) -- (-1,2.1);
	\draw [thick, purple] (0,-2) -- (-2,2.1);
	\draw [thick, purple] (0,-2) -- (-3,2.1);
	\draw [thick, purple] (0,-2) -- (0,2.1);

\end{tikzpicture}
\end{center}
\captionsetup{format=hang}
\caption{\small{The comoving observers in a Milne-like spacetime. They all emanate from the origin $\ms{O}$.}}\label{comoving figure in intro}
\end{figure}

It's interesting to understand the behavior of the comoving observers within the extended spacetime. Recall that the \emph{comoving observers} are the integral curves of $u = \pd_\tau$ and hence are given by the curves $\tau \mapsto (\tau, R_0, \theta_0, \phi_0)$ for various points $(R_0, \theta_0, \phi_0)$ on the hyperboloid. Physically, the comoving observers in an FLRW spacetime model the trajectories of the material particles which make up the galaxies, dust, etc. within the universe. In the $(t,r,\theta, \phi)$ coordinates, a comoving observer is given by $\tau \mapsto \big(t(\tau), r(\tau), \theta_0, \phi_0\big)$. By eq. (\ref{t and r def}), we have $t(\tau) = \coth(R_0)r(\tau)$. Thus, in the $(t,r,\theta, \phi)$ coordinates, the comoving observers are straight lines emanating from the origin $\mathcal{O}$. See figure \ref{comoving figure in intro}. This behavior can also be seen by noticing that the comoving observers have to be orthogonal to the hypersurfaces of constant $\tau$ which are the hyperboloids shown in figure \ref{milne universe and milne-like figure}. 

\medskip

Lastly, we note that the behavior illustrated in figure \ref{comoving figure in intro} is closely related to the notion of a \emph{Janus point}, see \cite{janus_point, janus_point_book}. For Milne-like spacetimes, the ``two-futures-one-past" scenario associated with a Janus point can be seen in\cite[figures 6 and 18]{Ling_coord_sing}.

\medskip

\subsection{The cosmological constant appears as an initial condition for Milne-like spacetimes}\label{Milne-like cosmo const sec}

As shown in \cite[Thm. 12.11]{ON}, FLRW spacetimes satisfy the Einstein equations with a perfect fluid $(u, \rho, p)$,
\begin{equation}
\text{Ric} - \half Rg\,=\, 8\pi T \,=\, 8\pi\big[(\rho + p)u_* \otimes u_* + pg\big],
\end{equation}
where $u_* = g(u,\cdot)$ is the one-form metrically equivalent to the vector field $u = \pd_\tau$.
We emphasize that for FLRW spacetimes, the energy density $\rho$ and pressure $p$ are purely geometrical quantities given by $\rho = \frac{1}{8\pi} G(u,u)$ and $p = \frac{1}{8\pi}G(e,e)$ where $e$ is any unit spacelike vector orthogonal to $u$ (its choice does not matter by isotropy). Here $G = \text{Ric} - \half Rg$ is the Einstein tensor which is related to $T$ via $G = 8\pi T$. To incorporate a cosmological constant $\Lambda$, we define $T_{\rm normal} = T + \frac{\Lambda}{8\pi}g$ so that the Einstein equations become
\begin{equation}
\text{Ric} - \half R g + \Lambda g \,=\, 8\pi T_{\rm normal}.
\end{equation}
Setting $\rho_{\rm normal} = T_{\rm normal}(u,u)$ and $p_{\rm normal} = T_{\rm normal}(e,e)$, we have 
\begin{equation}\rho_{\rm normal} \,=\, \rho - \rho_{\Lambda} \:\:\:\:  \text{ and } \:\:\:\: p_{\rm normal} \,=\, p - p_{\Lambda},
\end{equation}
where $\rho_{\Lambda} = \frac{\Lambda}{8\pi}$ and $p_{\Lambda} = -\frac{\Lambda}{8\pi}$. Note that 
\begin{equation}\label{coso const eq st} 
\rho_{\Lambda} \,=\, - p_{\Lambda}.
\end{equation} 
Eq. (\ref{coso const eq st}) is the \emph{equation of state} for a cosmological constant. 

For a $k = -1$ FLRW spacetime, the Friedmann equations \cite[Thm. 12.11]{ON} are given by

\begin{equation}\label{Friedmann eqs}
\frac{8\pi}{3}\rho(\tau) \,=\, \frac{a'(\tau)^2 - 1}{a(\tau)^2} \:\:\:\: \text{ and } \:\:\:\: -8\pi p(\tau) \,=\, \frac{2a''(\tau)a(\tau) + a'(\tau)^2 -1}{a(\tau)^2}.
\end{equation}
Now assume $(M,g)$ is Milne-like. For simplicity, assume that the scale factor is analytic at zero: $a(\tau) = \tau + \sum_2^\infty c_n\tau^n$. Taking the limit $\tau \to 0$ in (\ref{Friedmann eqs}), we find:
\begin{equation}\label{rho and p for Milne}
c_2 \,=\, 0 \:\:\:\: \Longrightarrow \:\:\:\:\rho(0) \,=\, -p(0) \,=\, \frac{3}{8\pi}(6c_3).
\end{equation}
Given eq. (\ref{coso const eq st}), the statement in (\ref{rho and p for Milne}) is what we mean by \emph{the cosmological constant appears as an initial condition for Milne-like spacetimes.} To obtain the same result under more relaxed assumptions on the scale factor, see \cite[Thm. 4.2]{Ling_coord_sing}. We generalize statement (\ref{rho and p for Milne}) in Theorems \ref{main} and \ref{main2} in the next section. Statement (\ref{rho and p for Milne}) won't hold for FLRW models which begin with a radiation dominated era (i.e. no inflation) since $\rho$ and $p$ diverge as $\tau \to 0$ \cite[exercise 12.14]{ON}.

Lastly, the scalar curvature for $(M,g)$ is given by
\begin{equation}\label{scsalar curv eq}
R(\tau) \,=\, 6\frac{a''(\tau)a(\tau) + a'(\tau)^2 -1}{a(\tau)^2}.
\end{equation}
Taking the limit $\tau \to 0$ in (\ref{scsalar curv eq}), we have
\begin{equation}\label{scalar curv = rho}
c_2 \,=\, 0 \:\:\:\: \Longrightarrow \:\:\:\: R(0) \,=\, 12(6c_3) \,=\, 32\pi\rho(0).
\end{equation}
We generalize statement (\ref{scalar curv = rho}) in Corollary \ref{cor 1} in the next section.

\medskip

\section{Main result}\label{main result}

\smallskip

In this section, we generalize the results of the previous section to spacetimes that share similar geometrical properties with Milne-like spacetimes but without any homogeneous or isotropic assumptions. Specifically, Theorems \ref{main} and \ref{main2} generalize statement (\ref{rho and p for Milne}) and Corollary \ref{cor 1} generalizes statement (\ref{scalar curv = rho}). We also deduce a statement about the Ricci curvature in Corollary \ref{cor 2}. 

Our definition of a spacetime $(M,g)$ will follow \cite{Ling_causal_theory}. In particular, the manifold $M$ is always assumed to be smooth. A \emph{smooth} spacetime is one where the metric $g$ is smooth, that is, its components $g_{\mu\nu} = g(\pd_\mu, \pd_\nu)$ are smooth functions with respect to any coordinates $(x^0, \dotsc, x^n)$.  A \emph{continuous} spacetime is one where the metric is continuous, that is, its components are continuous functions with respect to any coordinates. 


Let $(M,g)$ be a continuous spacetime. Our definition of timelike curves and the timelike future and past, $I^\pm$, will also follow \cite{Ling_causal_theory}. In particular, a \emph{future directed timelike curve} $\g \colon [a,b] \to M$ is a Lipschitz curve that's future directed timelike almost everywhere and satisfies $g(\g', \g') < -\e$ almost everywhere for some $\e > 0$. This class of timelike curves contains the class of piecewise $C^1$ timelike curves \cite[Prop. 2.4]{Ling_causal_theory}.

Let $(M,g)$ be a smooth spacetime. A continuous spacetime $(M_\ext, g_\ext)$ is said to be a \emph{continuous extension} of $(M,g)$ provided $M$ and $M_\ext$ have the same dimension, and there is an isometric embedding 
\[
(M,g) \,\hookrightarrow\, (M_\ext, g_\ext)
\]
preserving time orientations such that $M \subset M_\ext$ is a proper subset. Note that we are identifying $M$ with its image under the embedding.

Let $(M_\ext, g_\ext)$ be a continuous extension of a smooth spacetime $(M,g)$. The topological boundary of $M$ within $M_\ext$ is denoted by $\pd M = \ov{M} \setminus M$. A future directed timelike curve $\g \colon [a,b] \to M_\ext$ is called a \emph{future terminating timelike curve} for a point $p \in \pd M$ provided $\g(b) = p$ and $\g\big([a,b)\big) \subset M$. \emph{Past terminating} timelike curves are defined time-dually. The \emph{future} and \emph{past boundaries} of $M$ within $M_\ext$ are defined as  
\begin{align*}
\pd^+M \,&=\, \{p \in \pd M \mid \text{there is a future terminating timelike curve for $p$}\}\\
 \pd^-M \,&=\, \{p \in \pd M \mid \text{there is a past terminating timelike curve for $p$}\}.
\end{align*}
For example, $\pd^-M$ for a Milne-like spacetime coincides with the lightcone in figure \ref{milne universe and milne-like figure}.

A set in a spacetime is \emph{achronal} if no two points in the set can be joined by a future directed timelike curve. An important result we will use is the following lemma.\footnote{See \cite[Thm. 2.6]{GalLing_con} for a proof. The proof generalizes to the class of timelike curves considered in this paper since it only uses the openness of $I^\pm$ which follows from \cite[Thm. 2.12]{Ling_causal_theory}. Moreover, the ``topological hypersurface" part of the conclusion  follows from \cite[Thm. A.6]{Ling_causal_theory}.}

\medskip

\begin{lem}\label{future and past boundary lem}
If $\pd^+M = \emptyset$, then $\pd^-M$ is an achronal topological hypersurface.
\end{lem}

\medskip

For a Milne-like spacetime $(M,g)$, statement (\ref{rho and p for Milne}) implies that $\rho(\tau)$ and $p(\tau)$ extend continuously to $\tau = 0$ along each integral curve of $u$. We will use a slightly stronger version of these ``continuous extensions" which we make precise next.  

Suppose $(M_\ext, g_\ext)$ is a continuous extension of a smooth spacetime $(M,g)$ such that $M = I^+(\mc{O}, M_\ext)$ for some point $\mc{O} \in \pd^-M$. Let $f$ be a smooth function on $M$. We say $f$ \emph{extends continuously} to $M \cup \{\mc{O}\}$ provided there is a continuous function $\wt{f} \colon M \cup \{\mc{O}\} \to \R$ such that $\wt{f}|_M = f$. In this case, we call $\wt{f}$ the \emph{continuous extension} of $f$. The topology on $M \cup \{\mc{O}\}$ is the subspace topology inherited from $M_\ext$. In other words, $\wt{f}$ is continuous at $\mc{O}$ means that given any $\e > 0$, there is a neighborhood $U \subset M_\ext$ of $\mc{O}$ such that $|\wt{f}(\mc{O}) - \wt{f}(x)| < \e$ for all $x \in U \cap (M \cup \{\mc{O}\})$. 

Likewise, a smooth vector field $X$ on $M$ \emph{extends continuously} to $M \cup \{\mc{O}\}$ provided there is a coordinate neighborhood $U$ of $\mc{O}$ with coordinates $(x^0, \dotsc, x^n)$ such that each of the components $X^\mu$ in $X = X^\mu \pd_\mu$ extends continuously to $(U \cap M) \cup \{\mc{O}\}$. A similar definition applies to smooth tensors on $M$ by requiring each of its components to extend continuously. (This definition does not depend on the choice of coordinate system by the usual transformation law for tensor components.) For example, the metric tensor $g$ on $M$ extends continuously to $M \cup \{\mc{O}\}$ since $(M_\ext, g_\ext)$ is a continuous extension of $(M,g)$. For another example, suppose $T$ is a smooth tensor on $(M_\ext, g_\ext)$, then obviously the restriction, $T|_M$, extends continuously to $M \cup \{\mc{O}\}$ since it extends smoothly.

We are now ready to state our main result.

\medskip

\begin{thm}\label{main}
Let $(M_\ext, g_\ext)$ be a continuous extension of a smooth spacetime $(M,g)$ such that $M = I^+(\mc{O}, M_\ext)$ for some point $\mc{O} \in \pd^-M$. We make the following assumptions.

\begin{itemize}
\item[\emph{(a)}] $(M,g)$ solves the Einstein equations with a perfect fluid $(u, \rho, p)$. 

\item[\emph{(b)}] All the integral curves of $u$ have past endpoint $\mc{O}$ within $M_\ext$. Technical assumption: each of these extended curves are future directed timelike on any compact domain.

\item[\emph{(c)}] The Ricci tensor ${\rm Ric}$ of $(M,g)$, $\rho$, and $p$ extend continuously to $M \cup \{\mc{O}\}$.

\item[\emph{(d)}] $(M_\ext, g_\ext)$ is strongly causal at $\mc{O}$.
\end{itemize}
Then the continuous extensions of $\rho$ and $p$ satisfy $\wt{\rho} = -\wt{p}$ at $\mc{O}$.

\end{thm}

%

\medskip

\noindent\emph{Remarks.}

\begin{itemize}

\item[-] Recall that $\rho = -p$ is the equation of state for a cosmological constant. The conclusion of Theorem \ref{main} is that $\wt{\rho}(\mc{O}) = -\wt{p}(\mc{O})$; this is what we mean by the cosmological constant \emph{appears as an initial condition.} By continuity, we have $\rho \approx -p$ for points in $M$ near $\mc{O}$; this yields a ``quasi de Sitter" expansion which we elaborate more on in the next section.

\item[-] Note that  $M = I^+(\mc{O}, M_\ext)$ holds for Milne-like spacetimes; see figure \ref{milne universe and milne-like figure}. Assumption (b) mimics what happens in figure \ref{comoving figure in intro}. Hence the hypotheses in Theorem \ref{main} generalize what happens in a Milne-like spacetime but without the spatially isotropic assumption. The technical assumption in (b) says that if $\g \colon [0, b] \to M \cup \{\mc{O}\}$ is an integral curve of $u$ on $(0,b]$ with past endpoint $\g(0) = \mc{O}$, then $\g$ is a future directed timelike curve within $M_\ext$. Clearly $\g|_{[\e, b]}$ is a future directed timelike curve for any $\e > 0$. Since $g(\g', \g') = -1$ almost everywhere, requiring $\g$ to be future directed timelike amounts to $\g$ satisfying a Lipschitz condition. This would be satisfied, for example, if $\g$ was continuously differentiable at $\tau = 0$ (which holds for Milne-like spacetimes).

\item[-] Regarding assumption (c), let $(M,g)$ be a Milne-like spacetime with a scale factor that's analytic at zero: $a(\tau) = \tau + \sum_{2}^\infty c_n\tau^n$. If $c_2 \neq 0$, then it's easy to see from eq. (\ref{Friedmann eqs}) that $\rho$ and $p$ diverge as $\tau \to 0$. So our assumption that $\rho$ and $p$ extend continuously to $M \cup \{\mc{O}\}$ is similar to setting $c_2 = 0$ in statement (\ref{rho and p for Milne}). Moreover, if $c_2 = 0$ and $c_4 = 0$, then the Ricci tensor, $\text{Ric}$, of $(M,g)$ extends continuously to $M \cup \{\mc{O}\}$. (In fact $\text{Ric}$ extends continuously to $M \cup \pd^-M$). This follows from \cite[Lem. 3.5]{Ling_coord_sing} since $\text{Ric}$ can be written as a sum of products of the metric, its inverse, and their first and second derivatives along with the fact that the inverse metric is as regular as the metric.

\item[-] Regarding assumption (d), recall that $(M_\ext, g_\ext)$ is \emph{strongly causal} at $\mc{O}$ means that for any neighborhood $U$ of $\mc{O}$ there is a neighborhood $V \subset U$ of $\mc{O}$ such that $\g(a),\g(b) \in V$ implies $\g\big([a,b]\big) \subset U$ whenever $\g \colon [a,b] \to M_\ext$ is a future directed causal curve. This assumption holds for the usual continuous extensions of Milne-like spacetimes constructed in section \ref{Milne-like ext sec} since these constructions are conformal to (subsets of) Minkowski space which are strongly causal at every point.

\end{itemize}

\medskip

\noindent\underline{\emph{Proof of Theorem \emph{\ref{main}}}}.

\medskip
Before providing the details, we briefly outline the proof: If $\wt{\rho}(\mc{O}) \neq -\wt{p}(\mc{O})$, then assumptions (a) and (c) imply that the vector field $u$ extends continuously to $M \cup \{\mc{O}\}$. However, assumptions (b) and (d) along with $M = I^+(\mc{O}, M_\ext)$ imply that $M \cup \pd^-M$ locally looks like figure \ref{comoving figure in intro} near $\mc{O}$. It's evident from  figure \ref{comoving figure in intro} that $u$ does not extend continuously to $M \cup \{\mc{O}\}$, yielding a contradiction.

Seeking a contradiction, suppose $\wt{\rho} \neq - \wt{p}$ at $\mc{O}$. We show that this implies that $u$ extends continuously to $M \cup \{\mc{O}\}$. Since $(M,g)$ solves the Einstein equations with a perfect fluid, we have
\[
\text{Ric} - \half Rg\,=\, 8\pi T \,=\, 8\pi\big[(\rho + p)u_* \otimes u_* + pg\big]
\]
within $M$. Here $u_* = g(u, \cdot)$ is the one-form metrically equivalent to the vector field $u$. Since $\text{Ric}$ extends continuously to $M \cup \{\mc{O}\}$, so does the scalar curvature $R$ and hence so does $T$. 

Since $\wt{\rho}(\mc{O}) \neq -\wt{p}(\mc{O})$, there is a coordinate neighborhood $U \subset M_\ext$ of $\mc{O}$ such that $\wt{\rho} + \wt{p} \neq 0$ in  $U \cap (M \cup \{\mc{O}\})$. Then, within $U \cap M$, we have 
\[
u_* \otimes u_* \,=\, \frac{1}{\rho + p}(T - pg).
\]
The right-hand side of the above equality extends continuously to $(U \cap M) \cup \{\mc{O}\}$, hence so does the left-hand side. Let $S$ denote the continuous extension of $u_* \otimes u_*$ to $M \cup \{\mc{O}\}$. Let $(x^0, \dotsc, x^n)$ denote the coordinates on $U$. Let $S_{\mu\nu} = S(\pd_\mu, \pd_\nu)$. Then $S_{\mu\nu} = u_\mu u_\nu$ within $U \cap M$ where $u_\mu$ are the components of $u_*$. Define $\wt{u}_*$ on $M \cup \{\mc{O}\}$ via  $\wt{u}_*|_M = u_*$ and the extension 
\[
\wt{u}_\mu(\mc{O}) \,=\, \left\{
\begin{array}{ll}
      +\sqrt{S_{\mu\mu}(\mc{O})} & \text{ if } S_{\mu\mu}(\mc{O}) \neq 0 \text{ and } u_\mu(\mc{O}) > 0 \text{ near } \mc{O} \\
      -\sqrt{S_{\mu\mu}(\mc{O})} & \text{ if } S_{\mu\mu}(\mc{O}) \neq 0 \text{ and } u_\mu(\mc{O}) < 0 \text{ near } \mc{O}
      \\
      0 & \text{ if } S_{\mu\mu}(\mc{O}) = 0 
\end{array} 
\right.
\]
Then $\wt{u}_*$ is a continuous extension of $u_*$ to $M \cup \{\mc{O}\}$. Let $\wt{u}$ denote the vector field metrically equivalent to $\wt{u}_*$ (i.e. its components are given by $\wt{u}^\mu = g_\ext^{\mu\nu} \wt{u}_\nu$). Then $\wt{u}$ is a continuous extension of $u$ to $M \cup \{\mc{O}\}$. 

Since $g(u, u) = -1$ (by definition of a perfect fluid), continuity implies $g_\ext(\wt{u}, \wt{u}) = -1$ at $\mc{O}$. Using \cite[Lem. 2.9]{Ling_causal_theory} and applying the Gram-Schmidt orthogonalization process appropriately, for any $0 < \e <1$, we can assume that the coordinates $(x^0, \dotsc, x^n)$ on $U$ satisfy assumptions (1) - (6) below.

\begin{itemize}
\item[(1)] $\pd_0|_{\mc{O}} = \wt{u}(\mc{O})$,
\item[(2)] $x^0$ is a time function on $U$,
\item[(3)] $\wt{g}_{\mu\nu}(\mc{O}) = \eta_{\mu\nu}$ and $|\wt{g}_{\mu\nu}(x) - \eta_{\mu\nu}| < \e$ for all $x \in U$ where $\wt{g}_{\mu\nu} = g_\ext(\pd_\mu, \pd_\nu)$.
\end{itemize}
Here $\eta_{\mu\nu}$ are the usual components of the Minkowski metric with respect to the coordinates $(x^0, \dotsc, x^n)$. That is,  
\[
\eta \,=\, \eta_{\mu\nu}dx^\mu dx^\nu \,=\, -(dx^0)^2 + \delta_{ij}dx^idx^j.
\]
By choosing $U$ even smaller, we can also assume that
\begin{itemize}
\item[(4)] $\eta^\e(X,X) \leq 0 \,\Longrightarrow\, g_\ext(X,X) < 0$ for all nonzero $X \in T_pM_{\ext}$ whenever $p \in U$,
\end{itemize}
where $\eta^\e$ is the narrow Minkowskian metric on $U$ given by
\[
\eta^\e \,=\, -\frac{1-\e}{1+\e}(dx^0)^2 + \delta_{ij}dx^i dx^j \,=\, \eta + \frac{2\e}{1-\e}(dx^0)^2.
\]
Moreover, since $\wt{u}$ is a continuous extension of $u$ to $M \cup \{\mc{O}\}$, we can also assume that
\begin{itemize}
\item[(5)] $|\wt{u}^\mu (x) - \wt{u}^\mu(\mc{O})| < \frac{\e}{2}$ for all $x \in U \cap (M \cup \{\mc{O}\})$.
\end{itemize}
 Lastly, if $\phi \colon U \to \R^{n+1}$ denotes the coordinate map (i.e. $\phi = (x^0, \dotsc, x^n)$), then, by restricting the domain of $\phi$, we can assume that 
\begin{itemize}
\item[(6)] $\phi(U) = B_{2r}$ where $B_{2r} \subset \R^{n+1}$ is an open ball with radius $2r > 0$ (as measured by the Euclidean metric $\delta = \delta_{\mu\nu}dx^\mu dx^\nu$ on $U$) centered at the origin: $\phi(\mc{O}) = (0, \dotsc, 0)$.  
\end{itemize}

Choose $\e = \frac{3}{5}$. Then $\eta^\e$ has lightcones with `slope' $2$. Define the curve $c \colon [0, r] \to B_{2r}$ by $c(t) = (t, \frac{t}{2}, 0, \dotsc, 0)$. By (4), the curve $\phi^{-1}\circ c(t)$ is future directed timelike. Let $q = \phi^{-1} \circ c(r)$. Since $M = I^+(\mc{O}, M_\ext)$, it follows that $q \in M$. 
Let $\g\colon [0,b] \to M \cup \{\mc{O}\}$ denote the integral curve of $u$, i.e. $\g'(\tau) = u\circ \g(\tau)$ on $(0,b]$, with future endpoint $\g(b) = q$ and past endpoint $\g(0) = \mc{O}$. Note that $\tau$ is the proper time of $\g$.

\medskip
\medskip

\noindent{\bf Claim.} We can assume $\g\big([0,b]\big) \subset U$.

\medskip
\medskip

The claim follows by strong causality of $(M_\ext, g_\ext)$ at $\mc{O}$. To see this, note that strong causality implies that there is a neighborhood $V \subset U$ of $\mc{O}$ such that if $\g$ has endpoints in $V$, then the image of $\g$ is contained in $U$. Let $V' \subset V$ denote a neighborhood of $\mc{O}$ satisfying assumption (6) above. Then we work in $V'$ to construct the curve $\g$ in exactly the same way as in the paragraph above the claim. Then strong causality implies that the image of $\g$ is contained in $U$. This proves the claim.

By the claim and (2), we can reparameterize $\g$ by $x^0$. Let $\bar{\gamma} \colon [0,r] \to M \cup \{\mc{O}\}$ be the reparameterization of $\g$ by $x^0$. Then
\[
\bar{\g}(t) \,=\, \g \circ (x^0 \circ \g)^{-1}(t) \:\:\:\: \text{ where } \:\:\:\: x^0 \circ \g(\tau) \,=\, \int_0^\tau \frac{d(x^0 \circ \g)}{d\tau'}d\tau'.
\]
Note that  $\bar{\gamma}(0) = \mc{O}$ and $\bar{\gamma}(r) = q$. Since $\phi(q) = (r, \frac{r}{2}, 0, \dotsc, 0)$, the mean value theorem implies that there exists a $t_* \in (0,r)$ such that $(x^1 \circ \bar{\gamma})'(t_*) = \frac{1}{2}$. Set  $\gamma^\mu = x^\mu \circ \gamma$ and $\bar{\gamma}^\mu = x^\mu \circ \bar{\gamma}$. Using the fact that $\tau$ and $t = x^0 \circ \g$ are inverses of each other, the chain rule gives
\[
\frac{1}{2} \,=\, \frac{d \bar{\gamma}^{1}}{dt}(t_*) \,=\, \frac{d\g^1}{d\tau}\big(\tau(t_*)\big)\frac{d\tau}{dt}(t_*)\,=\, \frac{d\gamma^1/d\tau}{d\g^0 /d\tau}\big(\tau(t_*)\big) \,=\, \frac{u^1}{u^0}\big(\bar{\g}(t_*)\big). 
\]
However, by (1) and (5), we have
\[
\sup_{x \in U}\,\frac{u^1}{u^0}(x) \,\leq\, \frac{0 + \e/2}{1 - \e/2} \,=\, \frac{3}{7} \,<\, \frac{1}{2},
\]
which is a contradiction.
\qed

\medskip
\medskip

A careful inspection of the proof of Theorem \ref{main} reveals that assumption (d) is only used to prove the claim in the proof. The next theorem shows that one can replace assumption (d) with (d$'$). Essentially, (d$'$) says that $\pd^-M$ looks like figure \ref{milne universe and milne-like figure} at least locally near $\mc{O}$.

\medskip
\medskip

\begin{thm}\label{main2}
Let $(M_\ext, g_\ext)$ be a continuous extension of a smooth spacetime $(M,g)$ such that $M = I^+(\mc{O}, M_\ext)$ for some $\mc{O} \in \pd^-M$. Assume \emph{(a) - (c)} from Theorem \emph{\ref{main}} but replace assumption \emph{(d)} with 

\begin{itemize}
\item[\emph{(d$'$)}] For any neighborhood $U$ of $\mc{O}$, there is a neighborhood $V \subset U$ of $\mathcal{O}$ such that the past boundary of $M$ satisfies $\pd^-M \cap V \subset J^+(\mathcal{O}, V)$. 
\end{itemize}
Then the continuous extensions of $\rho$ and $p$ satisfy $\wt{\rho} = -\wt{p}$ at $\mc{O}$. 
\end{thm}

\proof
From the discussion above Theorem \ref{main2}, it suffices to show that the claim in the proof of Theorem \ref{main} holds. That is, we want to show that we could have chosen our neighborhood $U$ such that $\g\big([0,b]\big) \subset U$. 

Let $U'$ be a neighborhood of $\mc{O}$ satisfying (1)-(5) in the proof of Theorem \ref{main}. By assumption (d$'$), there is a neighborhood $V \subset U'$ of $\mc{O}$ such that $\pd^-M \cap V \subset J^+(\mc{O}, V)$. Let $U \subset U' \cap V$ be a neighborhood of $\mc{O}$ satisfying 
\begin{itemize}
\item[(6$'$)] $\phi(U) = (-2r, 2r) \times (-10r, 10r)^n$ for some $r > 0$ where $n + 1$ is the dimension of the spacetime. Assume that $U$ is still centered at the origin: $\phi(\mc{O}) = (0, \dotsc, 0)$.  
\end{itemize}

Again, choose $\e = 3/5$ and define the curve $c \colon [0,r] \to \phi(U)$ by $c(t) = (t, \frac{t}{2}, 0, \dotsc, 0)$. Let $q = \phi^{-1}\circ c(r)$. Again, we have $q \in M$, so let $\g \colon [0, b] \to M \cup \{\mc{O}\}$ denote the integral curve of $u$ with future endpoint $q$ and past endpoint $\mc{O}$. As remarked above, it suffices to show $\g \big([0,b] \big) \subset U$. 

Seeking a contradiction, suppose this is not the case. Define 
\[\tau_0 \,=\, \inf \{\tau \in [0,b] \mid \g\big((\tau,b]\big) \subset U\}.
\] 
 Then $\tau_0 > 0$ by assumption and $\g(\tau_0) \in \pd U$.  Since $\e = 3/5$ (and hence lightcones are contained within wider Minkowski lightcones with slope 1/2), applying \cite[Lem. 2.9 and 2.11]{Ling_causal_theory} shows that
\begin{itemize}
\item[(i)] $\lim_{\tau \to \tau_0} x^0 \circ \g(\t) \,=\, -2r.$
\end{itemize}
Since $\pd^-M \cap U \subset \pd^-M \cap V \subset J^+(\mc{O}, V) \subset J^+(\mc{O}, U')$, another application of \cite[Lem. 2.9 and 2.11]{Ling_causal_theory} gives
\begin{itemize}
\item[(ii)]  $x^0\big(\pd^-M \cap U) \subset [0, 2r)$. 
\end{itemize}

Since $M = I^+(\mc{O}, M_\ext)$, we have $\pd^+M = \emptyset$. Therefore, by Lemma \ref{future and past boundary lem}, $\pd^-M$ is an achronal topological hypersurface. Since it's a topological hypersurface, we can assume that it separates $U$ by shrinking $U$ if necessary. The separation is given by the following disjoint union 
\[
U \,=\, I^+(\pd^-M, U) \sqcup (\pd^-M \cap U) \sqcup \big(U \setminus \ov{I^+(\pd^-M, U)}\big).
\]
We have $q \in I^+(\pd^-M, U)$. By (i) and (ii), it follows that there must be some $\tau_* \in (\tau_0,b)$ such that $\g(\tau_*) \in \pd^-M$. However, this contradicts the achronality of $\pd^-M$  since $\g|_{[0,\tau_*]}$ is a future directed timelike curve with endpoints on $\pd^-M$.
\qed

\medskip
\medskip

\noindent\emph{Remark.} The assumption $M = I^+(\mc{O}, M_\ext)$ in Theorem \ref{main} can be replaced with the weaker assumption $I^+(\mc{O}, M_\ext) \subset M$; the proof remains unchanged. This is not true for Theorem \ref{main2} since $M \subset I^+(\mc{O}, M_\ext)$ was used to ensure that $\pd^+M = \emptyset$. However, we can make the weaker assumption  $I^+(\mc{O}, M_\ext) \subset M$ in Theorem \ref{main2} so long as we also assume conditions on $(M,g)$ that ensure $\pd^+M = \emptyset$. For example, if $(M,g)$ is future one connected and future divergent, then $\pd^+M = \emptyset$ \cite{SbierskiSchwarz1, GalLing_con}, or if $(M,g)$ is future timelike geodesically complete and globally hyperbolic, then $\pd^+M = \emptyset$ \cite{GLS}. In fact, future timelike geodesic completeness was shown to be sufficient \cite{Minguzzi_Suhr}.

\medskip
\medskip

\begin{cor}\label{cor 1}
Assume the hypotheses of either Theorem \emph{\ref{main}} or Theorem \emph{\ref{main2}}. Then the continuous extension of the scalar curvature satisfies 
\[
\wt{R}(\mc{O}) \,=\, 16\pi \frac{n+1}{n-1} \wt{\rho}(\mc{O}).
\] 
When the spacetime dimension is $n + 1 = 4$, we recover equation \emph{(\ref{scalar curv = rho})}.
\end{cor}

\proof
This follows from tracing the Einstein equations and using $\wt{\rho} = -\wt{p}$ at $\mc{O}$.
\qed

\medskip
\medskip

Corollary \ref{cor 2} shows that we can also get a statement about the Ricci curvature at $\mc{O}$. In some sense, it says that the spacetime ``begins Einstein." In fact Corollary \ref{cor 2} implies Corollary \ref{cor 1}; however, we have to work a bit harder to prove Corollary \ref{cor 2} since it relies on the technical assumption appearing in part (b) of Theorem \ref{main}.

%

\medskip
\begin{cor}\label{cor 2}
Assume the hypotheses of either Theorem \emph{\ref{main}} or Theorem \emph{\ref{main2}}.  Then the continuous extension of ${\rm Ric}$ to $M \cup \{\mc{O}\}$ satisfies
\[
\wt{{\rm Ric}} \,=\, \frac{16\pi \wt{\rho}}{n-1} \, g_\ext
\]
at $\mc{O}$. 
\end{cor}

\proof
Rewriting the Einstein equations, we have 
\[
\text{Ric} \,=\, 8\pi \big[(\rho + p)u_* \otimes u_* + pg\big] + \frac{8\pi}{n-1} \big[(\rho + p) - (n+1)p\big]g 
\]
within $M$. 

Let $U$ be a coordinate neighborhood of $\mc{O}$ with coordinates $(x^0, \dotsc, x^n)$. Write  $R_{\mu\nu} = \text{Ric}(\pd_\mu, \pd_\nu)$, and let $\wt{R}_{\mu\nu}$ denote their continuous extensions to $M \cup \{\mc{O}\}$. Consider a future directed timelike curve $\g\colon [0,b] \to M \cup \{\mc{O}\}$ where $\g$ is an integral curve of $u$ on $(0,b]$ with past endpoint $\g(0) = \mc{O}$. Setting $\g^\mu = x^\mu \circ \g$, we have
\[
R^{\mu\nu}\circ \g \,=\, 8\pi \left[(\rho + p)\frac{d\gamma^\mu}{d\tau}\frac{d\gamma^\nu}{d\tau} + p g^{\mu\nu} \right] + \frac{8\pi}{n-1}\big[(\rho + p) - (n+1)p\big]g^{\mu\nu}.
\]
Since $\g$ is future directed timelike, it satisfies a Lipschitz condition by definition. Therefore there is a constant $C$ such that 
\[
\left|\frac{d\gamma^\mu}{d\tau} \right| \,\leq\, C
\]
almost everywhere \cite[Prop. 2.2]{Ling_causal_theory}. 
Since $(\rho + p) \to 0$ along $\g(\tau)$ as $\tau \to 0$, the bound above implies that $(\rho + p)\frac{d\g^\mu}{d\tau}\frac{d\g^\nu}{d\tau} \to 0$ as $\tau \to 0$. Therefore
\[
\wt{R}^{\mu\nu}(\mc{O}) = -\frac{16\pi \wt{p}(\mc{O})}{n-1}\wt{g}^{\mu\nu}(\mc{O}) 
\]
where $\wt{g}^{\mu\nu}$ are the components of the inverse metric to $g_\ext$. The result follows.
\qed

%

\medskip
\medskip

In this section, we have always assumed that $\text{Ric}$ extends continuously to $M \cup \{\mc{O}\}$. Finding sufficient conditions on the perfect fluid $(u,\rho,p)$ for when this happens is perhaps an interesting question, but this will not be explored here. 

\section{Some remarks on inflationary scenarios}\label{inflationary section}

In this section, we show how the results from the previous section can be used to imply inflationary scenarios for spacetimes without the homogeneous and isotropic assumptions associated with FLRW spacetimes. The idea is this: the cosmological constant appearing as an initial condition yields a ``quasi de Sitter" expansion in the early universe. For FLRW spacetimes, this is seen via Friedmann's second equation. (We show this in statement (\ref{2nd friedmann eq}) below.) For the nonhomogeneous spacetimes considered in the previous section, we will use the Raychaudhuri equation, which is a generalization of Friedmann's second equation, to obtain an inflationary expansion.

\newpage

An inflationary era is characterized by an accelerated expansion, $a''(\tau) > 0$, right after the big bang but before the radiation dominated era. It's speculated to occur since it solves certain problems in cosmology (e.g. the horizon and flatness problems) and predicts that the spectrum of density perturbations is scale-invariant. For a nice introduction on inflationary theory, see \cite{Liddle}; for a more thorough account, see \cite{WeinbergCos}. The significance of $a''(\tau) > 0$ is that it violates the \emph{strong energy condition} which holds for all known physical matter models, e.g. dust and radiation. (It's also the energy condition appearing in Hawking's cosmological singularity theorems.) Therefore, if the energy-momentum tensor was dominated by radiation in the early universe immediately after the big bang, then an inflationary era cannot occur. Some other matter model, which violates the strong energy condition, must be used to generate an inflationary era.

 To account for an inflationary era, one normally introduces an ``inflaton" scalar field $\phi$ in a slow-roll potential. If the energy-momentum tensor was dominated by the scalar field, then the slow-roll potential implies $a''(\tau) > 0$. This is 
 \emph{not} our approach. Instead, we obtain an inflationary era from the \emph{geometry} of the spacetimes considered in the previous section. The geometry of these spacetimes, encoded in the assumptions of Theorems \ref{main}/\ref{main2}, imply that the cosmological constant appears as an initial condition which implies a ``quasi de Sitter" expansion. We show this next.

From eq. (\ref{Friedmann eqs}), we obtain Friedmann's second equation
\begin{equation}\label{2nd friedmann eq}
\frac{a''(\tau)}{a(\tau)} \,=\, -\frac{4\pi}{3}\big(\rho(\tau) + 3p(\tau)\big).
\end{equation}
Consider a Milne-like spacetime with $a(\tau) = \tau + \sum_2^\infty c_n\tau^n$ near $\tau = 0$. By statement (\ref{rho and p for Milne}) and eq. (\ref{2nd friedmann eq}), we see that
\begin{equation}\label{inflationary scenario eq}
c_2 \,=\, 0 \quad \Longrightarrow \quad \rho(0) \,=\, -p(0) \quad \Longrightarrow \quad a''(\tau) \,>\, 0 
\end{equation}
for $\tau$ near $\tau = 0$ provided $\rho(0) > 0$. Hence we see that the assumptions $c_2 = 0$ and $\rho(0) > 0$ yield an inflationary era. 

Next we generalize statement (\ref{inflationary scenario eq}) to the nonhomogeneous spacetimes considered in the previous section. First, we identify a geometric quantity for $a''$. Let $\{e_0, e_1, e_2, e_3\}$ be an orthonormal frame for an FLRW spacetime with $e_0 = u = \pd_\tau$. Using $\langle \cdot, \cdot \rangle$ to denote the metric $g(\cdot, \cdot)$, we have
\[
\text{div}(u) \,=\, -\langle \nabla_{e_0} u, e_0 \rangle + \sum_{i = 1}^3 \langle \nabla_{e_i} u, e_i \rangle \,=\, \sum_{i = 1}^3 \langle \nabla_{e_i} u, e_i \rangle \,=\, 3\frac{a'}{a}.
\]
Set $H = \frac{1}{3}\text{div}(u) = a'/a$. Then the equation $a''/a = (a'/a)' + (a'/a)^2$ becomes
\begin{equation}\label{a'' eq} 
a''/a \,=\, H' + H^2.
\end{equation}

The right hand side of eq. (\ref{a'' eq}) will be our geometrical substitute for $a''$. For FLRW spacetimes, $u$ is hypersurface orthogonal and so $H$ coincides with the mean curvature, $\frac{1}{3}\text{tr}(K)$, of the constant $\tau$-slices  where $K$ is the second fundamental form of the slice given by $K(X,Y) = \langle \nabla_X u, Y\rangle$.\footnote{Our convention for the mean curvature $H$, which includes the 1/3 factor in front of $\text{tr}(K)$, coincides with the Hubble parameter, $a'/a$, which is also denoted by $H$ in the physics literature.}

Now let $(M,g)$ be any smooth spacetime, and let $u$ be a smooth future directed timelike vector field on $M$ normalized to $\langle u,u\rangle = -1$. For simplicity, assume $\text{dim}(M) = 4$. Define $H = \frac{1}{3}\text{div}(u)$. Letting $\tau$ denote the proper time of the flow lines of $u$, the Raychaudhuri equation \cite[eq. (4.26)]{HE} gives
\begin{equation}\label{Ray eq}
3\left(\frac{d H}{d\tau} + H^2\right)\,=\, -\text{Ric}(u,u) + 2\omega^2 - 2\sigma^2 + \text{div}(\nabla_u u),
\end{equation}
where $2\omega^2 = \omega_{ij}\omega^{ij} \geq 0$ and $2\s^2 =\s_{ij}\s^{ij}\geq 0$. Here $\omega$ and $\sigma$ are the \emph{vorticity} and \emph{shear} scalars, which are completely determined by vectors spanning the orthogonal complement $u^\perp$, see                                                                       \cite[ch. 7 and 12]{Frankel_grav}. When $u$ is hypersurface orthogonal, the vorticity scalar vanishes and $H$ coincides with the mean curvature of the hypersurfaces.

Following \cite{Ellis, Ellis_Elst}, we define an \emph{average length scale} $\mathfrak{a}(\tau)$ along the flow lines of $u$ via $\mathfrak{a}'/\mathfrak{a} = H$ where the prime $'$ denotes a derivative with respect to the proper time $\tau$ of the flow lines. With this definition, we have $\mathfrak{a}''/\mathfrak{a} = H' + H^2$ which generalizes eq. (\ref{a'' eq}). For FLRW spacetimes, the average length scale $\mathfrak{a}(\tau)$ coincides with the scale factor $a(\tau)$.

Consider the setting of the previous section. Assume the hypotheses of either Theorem \ref{main} or Theorem \ref{main2}. By Corollary \ref{cor 2}, for points near $\mc{O}$, eq. (\ref{Ray eq}) gives
\begin{equation}\label{a'' gen eq}
3\frac{\mathfrak{a}''}{\mathfrak{a}} \,\approx\,\, 8\pi \wt{\rho}(\mc{O}) + 2\omega^2 -2\s^2 + \text{div}(\nabla_u u).
\end{equation}
Eq. (\ref{a'' gen eq}) generalizes eq. (\ref{2nd friedmann eq}). If $2\s^2 - \text{div}(\nabla_u u)$ is sufficiently less than $8\pi \wt{\rho}(\mc{O})$ for points close to $\mc{O}$, then eq. (\ref{a'' gen eq}) shows that $\mathfrak{a}'' > 0$. Since $\mathfrak{a}''/\mathfrak{a} = H' + H^2$, eq. (\ref{a'' eq}) shows that we can interpret $\mathfrak{a}'' > 0$ as an analogue for an inflationary era in this nonhomogeneous setting.

If $u$ is a geodesic vector field (which is the case for FLRW spacetimes), then $\nabla_u u = 0$ and so we only require that $2\s^2$ is sufficiently less than $8\pi \wt{\rho}(\mc{O})$ to obtain $\mathfrak{a}'' > 0$. Recall that $2\s^2$ measures the rate of shear of the flow; it's  zero for FLRW spacetimes and, in fact, zero for any fluid flow with uniform expansion. In this sense, assuming $2\s^2$ is sufficiently small can be thought of as a substitute for the spatial isotropy associated with FLRW spacetimes.

\medskip

\section*{Acknowledgments} 
The author gratefully acknowledges being supported by the Harold H. Martin Postdoctoral Fellowship at Rutgers University. He thanks Greg Galloway for many helpful comments and pointing out references \cite{Ellis, Ellis_Elst}. He thanks two anonymous reviewers who greatly improved the quality of the paper. Lastly, he thanks the organizers of \emph{Singularity theorems, causality, and all that; a tribute to Roger Penrose} for putting together a stimulating conference.

\medskip
\medskip
\medskip

\noindent {\bf Data availability statement}
\medskip

 \noindent This manuscript has no associated data.

\bibliographystyle{amsplain}

\end{document}